# Experimental Realization of Two-Dimensional Boron Sheets

Baojie Feng[1], Jin Zhang[1], Qing Zhong[1], Wenbin Li[1], Shuai Li[1], Hui Li[1*], Peng Cheng[1], Sheng Meng[1,2], Lan Chen[1*] and Kehui Wu[1,2*]

[1]*Institute of Physics, Chinese Academy of Sciences, Beijing 100190, China*

[2] *Collaborative Innovation Center of Quantum Matter, Beijing 100871, China*

*huili8@iphy.ac.cn, lchen@iphy.ac.cn, khwu@iphy.ac.cn

**Boron is the fifth element in the periodic table and possesses rich chemistry second only to carbon [1-3]. A striking feature of boron is that $B_{12}$ icosahedral cages occur as the building blocks in bulk boron and many boron compounds [2,3]. This is in contrast to its neighboring element, carbon, which prefers 2D layered structure (graphite) in its bulk form. On the other hand, boron clusters of medium size have been predicted to be planar or quasi-planar, such as $B_{12}^{+}$, $B_{13}^{+}$, $B_{19}^{-}$, $B_{36}$, and so on [4-10]. This is also in contrast to carbon clusters which exhibit various cage structures (fullerenes). Therefore, boron and carbon can be viewed as a set of complementary chemical systems in their bulk and cluster structures. Now, with the boom of graphene [11, 12], an intriguing question is that whether boron can also form a monoatomic-layer 2D sheet structure? Here, we report the first successful experimental realization of 2D boron sheets. We have revealed two types of boron sheet structures, corresponding to a triangular boron lattice with different arrangements of the hexagonal holes. Moreover, our boron sheets were found to be relatively stable against oxidization, and interacts only weekly with the substrate. The realization of such a long expected 2D boron sheet could open a door toward boron electronics, in analogous to the carbon electronics based on graphene.**

Graphene is the building block of various carbon nanostructures like fullerenes and nanotubes [11, 12]. It also provides a playground to build quantum electronics due to its fascinating Dirac-like electron properties. Boron has the same short covalent radius and flexibility to adopt $sp^2$ hybridization as carbon, which would favor the formation of various low-dimensional allotropes, such as boron nanotubes, fullerenes, and 2D boron sheets [13-15]. However, to date, a real 2D boron sheet has not yet been obtained. Theoretically, 2D boron sheets also exhibit attractive electronic properties. For example, bulk boron is a semiconductor, while 2D boron sheet would be metallic [16-20]. In fact, compound $MgB_2$, which is composed of alternating boron layers and Mg layers, exhibits high $T_c$ superconductivity at ambient pressure [21]. Even some 2D structures of boron were predicted to host Dirac fermions recently [22]. In fact, there have been tremendous theoretically works investigating possible 2D boron sheet structures [16-20]. A large variety of planar boron structures with competitive cohesive energy, such as ±-sheet [16,17], ²-sheet [16,17], and Ç-sheets [19,20], named by the connectivity of boron, have been predicted. However, experimental realization of 2D boron sheets is still a challenge. Recently it had been suggested that monolayer boron sheets can form on metal substrates, such as Cu(111) [23], Ag(111) and Au(111) [24], due to the stabilization of $sp^2$ hybridization by metal passivation. Following this idea, we employed molecular beam epitaxy (MBE) to grow boron sheets on Ag(111) surface in ultrahigh vacuum (UHV) chamber.

Boron was grown on a single crystal Ag(111) surface by direct evaporation of a pure boron source. When the substrate temperature during growth was below 500 K, we found only clusters or disordered structures on the surface (Fig. S1(a) in the supplemental materials (SM)). When the temperature was raised to about 570 K, monolayer islands with a perfectly ordered structure form on the surface, as shown in Fig.1(a). The STM image with high contrast (Fig.1(c)) shows characteristic feature of parallel stripes on island surface in the [$\bar{1}10$] direction of Ag(111). The periodicity of the stripes is 1.5 nm, as indicated by the solid lines in Fig.1(c). The STM image with high resolution shown in Fig.1(e) reveals ordered, parallel rows of protrusions along [$\bar{1}\bar{1}2$] direction of the Ag(111) substrate (horizontal direction in Fig.1(e)). The distance between nearest-neighbor protrusions is 3.0 Å along the rows and 5.0 Å across the rows, respectively (marked by the black rectangle). The 1.5 nm stripes observed in large scale images (black lines in Fig.1(c)) correspond to the slightly brighter protrusions which are aligned in the direction perpendicular to the rows, as marked by the lines in Fig. 1(e). In the following we label this phase as S1 phase.

By annealing the sample to 650 K, we observed the transition of S1 phase to another ordered structure (labeled as S2), as shown in Fig.1(b). The S2 phase usually coexists with S1 phase in the temperature range from 650 K to 800 K. At higher temperatures most areas of the surface will be transformed to S2 phase. On the other hand, S2 phase can also be obtained by directly growing B on Ag(111) with substrate temperature around 680 K (Fig. S1(c)). The high-resolution STM image (Fig. 1(d) and (f)) indicates that S2 phase also consists of parallel rows of protrusions in the [$\bar{1}\bar{1}2$] direction of Ag(111) surface. The distance between nearest-neighbor protrusions is 3.0 Å along the rows, and the inter-row distance is 4.3 Å. Another obvious feature is that the protrusions along the rows are divided into sections with alternative brighter and darker protrusions. Each section involves five protrusions, so the periodicity along the row is 3.0 Å ×5=1.5 nm.

As boron has very low solubility in bulk Ag, we first assume that these two structures are both pure boron sheets. Compared with a large number of freestanding planar boron monolayer models in literature, we find that the $\beta_{12}$-sheet structure [19,20] agrees with S1 phase very well. The $\beta_{12}$-sheet model is characterized by vacancy chains separated by hexagonal boron rows. Its unit cell is rectangular with lattice constants of 3.0 Å and 5.0 Å in two directions, respectively. To confirm this model, we performed first-principles calculations on the $\beta_{12}$-sheet involving the Ag(111) substrate. The $\beta_{12}$-sheet is placed on Ag(111) with the boron rows in the [$\bar{1}\bar{1}2$] direction of the Ag(111) surface, as shown in Fig. 2(a). After structural relaxation, the global structure of the $²_{12}$-sheet remains planar. It is noted that, due to the lattice mismatch, five times the lattice constant of the boron sheet along the rows (3.0 Å ×5=15.0 Å) fits well with three times the period of Ag(111) (3×2.9 Å×·3=15.06 Å). This means that a Moiré pattern with 1.5 nm periodicity will form along the boron row direction, which explains the 1.5 nm stripes observed in STM image (Fig. 1(e)) well. The simulated STM image of $²_{12}$-sheet on Ag(111) is shown in Fig.2(c), which features both the rectangular lattice and the parallel striped patterns, in accordance with the STM images.

On the other hand, the S2 phase most likely corresponds to the Ç₃-sheet model in literature [20], as shown in Fig. 2(b). The structure of Ç₃-sheet consists of similar, but narrower zig-zag boron rows. This can explain the observed slightly smaller inter-row distance in the S2 phase (4.3 Å) as compared with that of S1 phases (5.0 Å). Our first-principles calculation also suggests good commensuration between Ç₃-sheet and Ag(111), and the structure remains planar after relaxation, as shown in Fig.2(b). Similarly, the simulated STM image of Ç₃-sheet (Fig.2(d)) shows zig-zag rows and alternative bright-dark protrusions along the rows, agreeing perfectly with our STM observations.

Because the atomic structures of $²_{12}$-sheet and Ç₃-sheet on Ag(111) remain planar without obvious vertical undulation, the 1.5 nm periodicity along the vacancy chains in S1 phase and the dark-bright alteration in S2 phase should correspond to the modulation of the electron density on the boron sheets. To address this question, we calculated the atomic charges of these two phases based on the Bader analysis (Fig. S9 in SM). Inhomogeneous charge distribution is found along boron rows in both S1 and S2 phases. Such inhomogeneous charge distribution should come from the commensuration between the lattices of boron sheet and Ag(111), which results in different adsorption sites of B atoms on the Ag(111) substrate, especially along the boron row direction.

The validity of the above models has been confirmed experimentally, by measuring the atomic density of boron in the two phases. To precisely calibrate the boron coverage, we used the well-known Si(111)-B-$\sqrt{3}\times\sqrt{3}$ structure as a reference. Boron atoms were deposited subsequently onto the Ag(111) surface and a clean Si(111)-7×7 surface in the same deposition cycle, thus governing exactly the same flux and same position of samples. The B/Si(111) sample was then annealed to 1000 K to obtain a Si(111)-B-$\sqrt{3}\times\sqrt{3}$ surface. Each B atoms on Si(111) can be easily counted based on the STM images and thus the flux of B was precisely determined. The total atomic density of B grown on Ag(111) were determined by the calibrated B flux on Si(111) and the deposition time. Through counting the area ratio of S1 on Ag(111) in STM images, the atomic density of B in S1 was obtained. We have performed experiments on samples with different B coverages, and the calibrated atomic density of B in S1 phase is 33.6±2.0 nm$^{-2}$, identical to that of $\beta_{12}$-sheet model (34.48 nm$^{-2}$) within a small error range. The details of the above data and analysis are included in SM. On the other hand, when we annealed the S1 surface and turned it to the S2 phase, we did not observe a significant change of the total area of the islands, indicating that the boron density in the two structures are approximately the same. This is consistent with the fact that the $\chi_3$-sheet model has similar boron density (31.3 nm$^{-2}$) as that of the $\beta_{12}$-sheet. The above experimental facts confirm the validity of our structure models, and rule out the possibility of B-Ag alloying.

Increasing the boron coverage, the 2D boron sheets can extend their size until they percolate to cover almost the entire surface (Fig. S2 in SM). It should be noted that when the boron coverage is close to 1 monolayer (ML), pronounced amount of 3D clusters form. Due to the formation of the 3D boron clusters it is difficult to obtain high quality multilayer boron films. We suggest that this is because the Ag-B interface interaction is indispensable for stabilizing boron atoms in a 2D form. When the boron coverage exceeds 1ML, the interface interaction will saturate, resulting in spontaneous formation of 3D boron clusters

The chemical bonding in our 2D boron sheets on Ag(111) have been investigated by *ex-situ* X-ray photoelectron spectroscopy (XPS). In Fig.3 we show the XPS of a typical sample of pure S1 phase with coverage of 0.7 ML. The survey of XPS shown in Fig.3(a) contains B 1s signal, as well as C 1s and O 1s peaks which are due to air contamination during sample transfer in ambient condition. Fig.3(b) shows three peaks in the B 1s signal: 191.5 eV, 188.3 eV, and 187.2 eV, indicating that there are three types of boron atoms with different chemical environments. As the binding energy of B 1s peak in bulk boron is about 189-190 eV [25], the two low binding energy peaks (188.3 eV and 187.2 eV) which are slightly red shifted compared to the bulk value, are most likely originated from the B-B bond in the pristine 2D boron sheets. On the other hand, the higher energy peak (191.5 eV) implies oxidation of boron during exposure of samples to air, in accordance with the oxidized B peaks reported previously [26]. Importantly, the measured area ratio of the B-O peak to B-B peak is about 0.26, indicating that most boron atoms remain intact. We suggest that the oxidized boron atoms may come mostly from the edges of the sheets where unsaturated boron bonds exist, while the body of the sheets remains intact.

To explain these two B-B peaks, we calculated the atomic charges of 2D boron (S1 and S2 phases). The results are shown in Fig.S9 in SM. Interestingly, the B atoms in the boron sheets can be classified into two groups: the B atoms in the center of the hexagonal boron rows are negatively charged, and outer B atoms are positively charged. It means that the B atoms in 2D boron have two different chemical environments, agreeing well with the two B-B peaks observed in XPS. On the other hand, the XPS on another sample with B coverage about 1 ML shows similarly three B 1s peaks (Fig. 3(c)), but the area ratio of the B-O peak to the B-B peak increases to about 0.56. As the major difference between the low coverage and high coverage samples is the appearance of 3D boron clusters at high coverage (Fig. S2(c)), we suggest that the increased B-O peak mainly comes from the oxidation of 3D boron clusters.

Therefore, the 2D boron sheet is more inert to oxidation, as compared with the 3D boron cluster. Such chemical stability of 2D boron sheets is promising for future device applications.

Previously, a large number of 2D boron sheet structures have been theoretically predicted with competitive formation energies. However, among these models, neither $\beta_{12}$ nor $\chi_3$ corresponds to the global energy minimum [20]. To understand why we only observe these two structures, we calculated the formation energies of the S1 and S2 structures, as shown in Table I. The formation energy of freestanding S1 is 0.04 eV higher than S2, indicating a higher stability of isolated S1 over S2. Nevertheless, the adhesion energy of S2 (0.16 eV/atom) on Ag(111) is larger than that of S1 (0.09 eV/atom). Thus, the total formation energy for S1 (6.32 eV/atom) is actually slightly smaller than that of S2 (6.35 eV/atom). This means S2 is thermodynamically more stable when it adsorbs on Ag(111), which agrees with the higher thermal stability of S2 phase in our experiment. Therefore, the B-Ag interaction should play an important role in the formation of 2D B structures on Ag(111) substrate, resulting in optimal structures not necessarily corresponding to a global energy minimum in vacuum. Secondly, the perfect commensuration of the lattice parameters of $\beta_{12}$-sheet and $\chi_3$-sheet with that of Ag (111) can efficiently lower the strain, favoring their formation on Ag(111). Finally, 2D boron structures with hexagonal holes in the triangular lattice can be described by the holes density $\eta$, which is defined as the ratio of number of hexagonal holes to the number of B atoms in the original triangular lattice. The $\beta_{12}$-sheet and $\chi_3$-sheet have very close hole density (1/6 and 1/5) [20], which explains the easy transition of the two structures by annealing.

The fact that both the $\beta_{12}$-sheet and $\chi_3$-sheet maintain their isolated planar structures after adsorption on Ag(111) implies the interactions between the boron sheets and Ag(111) surface are not strong, in agreement with previous theoretical results [24]. Indeed, the calculated adhesion energy between S1 and Ag(111) is 0.03 eV/Å$^2$, in the same order as the binding energy of graphite (0.019 eV/Å$^2$) [27] and monolayer graphene on Cu(111) (0.022 eV/Å$^2$) [28]. This implies the S1 boron sheet may be separated from the substrate, in analogous to graphene, as also suggested by previous works [24]. The slightly larger adhesive energy (~0.05 eV/Å$^2$) for S2 on Ag(111) indicates S2 is more strongly bonded to the substrate, but still not difficult to detach. According to our first-principles calculations, the weak interfacial interaction is also reflected by the large distance between boron sheets and substrate (~2.4 Å), as well as very tiny charge transfer from Ag(111) to boron sheet (~0.03 e per B atom). Combined with the XPS results which indicate that boron sheets are hard to oxidize, it may be possible to obtain freestanding 2D boron sheets, with merits of transport and other device applications in the future.

Fig.4 shows the calculated band structures of the two types of boron sheets. Both boron structures are metallic, in accordance with the band structures in gas phase in previous calculations [19,20] and our DFT results (Fig S10 in SM). The scanning tunneling spectroscopy (STS) on 2D boron surface (Fig. S5(b) and (c)). Fig. S5(b) shows a prominent peak at -2.5 V (the STS of S1 and S2 are similar). The LDOS at bias range from -2V to +2V is relatively low. But the STS took at lower bias voltage (Fig. S5(c)) shows a significant density of states (DOS) around Fermi level, which proves that the 2D boron sheets are metallic. We also calculated the projected density of states (PDOS) of 2D boron (Fig. S6), which gives a high p-band peak at -2.0 eV (-2.5eV) for S1 (S2) phase and the DOS distribution around Fermi level, in good agreement with the experimental STS.

Finally, although the B-Ag interaction is found to be weak for our 2D boron sheet, intensive edge electronic states are evidenced by the brighter edges of 2D boron islands observed in STM images (Fig. 1(a) and (b)). The STS (Fig. S5(b) and (c)) on edges show additional shoulders around -3.0 eV and -0.2 eV, and the dI/dV maps at -3.2 eV and -0.2 eV (Fig. S5(e) and (g)) show most prominent bright edges, indicating the existence of edge states. Our first-principles calculations on boron nanoribbons (the details are shown in SM) indicate that the boron sheet are bonded to Ag(111) substrate predominately through their edges, where the electron density is much higher than inner part of the boron sheet. Such edge interaction, as well as the resulting edge state, may play important roles in the future electronic applications of 2D boron nanoribbons.

In summary, our experiments and first-principles calculations have confirmed the formation of two types of 2D boron structures on Ag(111): $\beta_{12}$-sheet and $\chi_3$-sheet. The boron sheets are quite inert to oxidation and interact only weakly with the Ag(111) substrate. Our results pave the way to exploring boron-based microelectronic device applications. Besides Ag(111), other substrates, such as Au(111), Cu(111) and metal borides [23,24], might also be good platform for the synthesis of monolayer boron sheets. Substrates with different interactions with boron may produce 2D boron sheets with different structures. The novel physical or chemical properties of boron sheets, such as massless Dirac fermions [22], novel reconstruction geometries [29], may be realized in the 2D boron sheets with special structure, which remains to be investigated in the future.

**Method:**

**Experiments:** The experiments were performed in a UHV chamber combing a MBE system and a low temperature (4.5 K) scanning tunneling microscope (STM) with base pressure of $2\times10^{-11}$ Torr. The samples were grown in the MBE chamber. Single crystal Ag(111) was cleaned by repeated $Ar^+$ ion sputtering and annealing cycles. Pure boron (99.9999%) was evaporated from an e-beam evaporator onto the clean Ag(111) substrate while keeping the substrate at appropriate temperatures. The pressure during boron growth was better than $6\times10^{-11}$ Torr. After growth, the sample was transferred to the STM chamber without breaking the vacuum. All the STM images and STS were taken at 78 K and the bias voltages were defined as the tip bias with respect to the sample. The X-ray photoelectron spectroscopy (XPS) experiments were performed *ex-situ* by taking out the as-prepared sample to air, and transferring into the XPS system for measurements (KRATOS, AXIS Untra DLD spectrometer with a Al K$_\alpha$ X-ray source, hν=1486.6 eV). The XPS instrumental resolution is 0.48 eV (determined from the Ag 3d$_{5/2}$ peak). The binding energy and Fermi level were calibrated by measurements on pure Au, Ag and Cu surfaces. XPS peaks were fitted with a mixture of Gaussian and Lorentzian functions.

**Calculations:** The density functional theory (DFT) calculations were performed using a projector-augmented wave (PAW) pseudopotential in conjunction with the Perdew-Burke-Ernzerhof (PBE) [30] function and plane-wave basis set with energy cutoff at 400 eV. For S1, the calculation cell contains a boron film on a five-layer 3×3 √3 Ag(111) surface. The surface Brillouin zone was sampled by 3×3×1 Monkhorst-Pack k-mesh. For S2, the Boron film was positioned on fiver-layer 3×3 Ag(111) surface. The Brillouin zone was sampled by 5×5×1 Monkhorst-Pack k-mesh. As PBE usually overestimates chemical bond length, the lattice constant of Cu(111) employed in the calculations was 3% larger than experimental value, and the vacuum region of about 15 Å was applied. All the structures were fully relaxed until the force on each atom was less than 0.05 eV/Å, and the bottom two layers of Ag atoms were fixed. The simulated STM images were obtained using constant current mode based on calculated electron densities. All the calculations were performed with Vienna Ab initio Simulation Package (VASP) [31].

**Acknowledgements**

This work was supported by the MOST of China (Grants No. 2012CB921703, 2013CB921702, 2013CBA01600), and the NSF of China (Grants No. 11334011, 11322431, 11174344, 91121003), and the Strategic Priority Research Program of the Chinese Academy of Sciences, Grant No. XDB07020100.

**Author Contributions**

K. W. and L. C. designed the experiments; B. F., L. C. Q. Z., W. L. and S. L. performed experiments as well as data analysis under the supervision of K. W.; J. Z. H. L. and S. M. performed the DFT calculations; B. F., L. C. and K. W. wrote the manuscript with contribution from all the authors; all co-authors contributed to data analyses and discussions.

Table I: **Formation energies for free-standing and epitaxial S1 and S2 boron sheets.** $E_{FB}$ is the formation energy per atom for free-standing B sheet; $E_{EB}$ is the formation energy per B atom for epitaxial B sheet; ″$E_1$ is adhesion energy of boron sheet per B atoms; ″$E_2$ is adhesion energy of boron sheet per unit area.

|  | S1 | S2 |
|---|---|---|
| $E_{FB}$ (eV/atom) | 6.23 | 6.19 |
| $E_{EB}$ (eV/atom) | 6.32 | 6.35 |
| ″$E_1$ (eV/atom) | 0.09 | 0.16 |
| ″$E_2$ (eV/Å$^2$) | 0.03 | 0.05 |

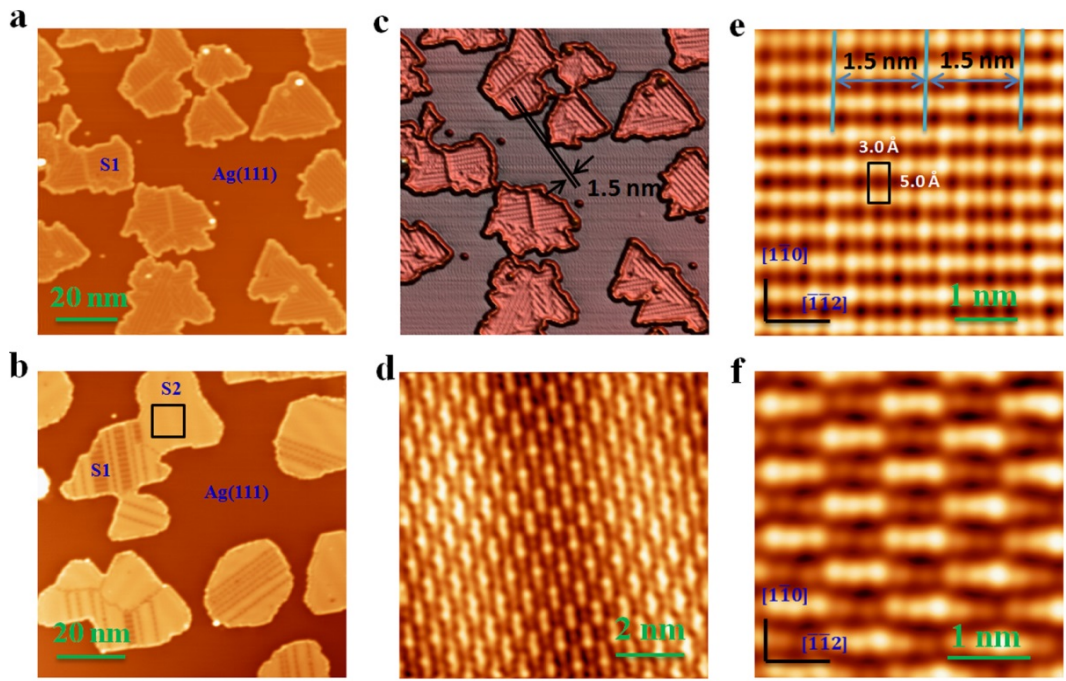

Fig. 1. **Formation of 2D boron sheets on Ag(111).** (a,b) STM topographic images of boron structures on Ag(111). The two different structures are labeled as S1 and S2; (b) is the result of annealing the surface in (a) to 650 K. Most boron islands are transferred to S2 phase but the S1 phase still remains in small parts of the islands. (c) 3D version of (a) in which the stripes with 1.5 nm distance can be resolved more clearly. (d) STM image obtained on the area marked by the black rectangle in (b). (e) High resolution STM images of S1 phases. (f) High resolution STM image of S2 phase, which is zoomed from (d). It is noted that the orientation of image is rotated to compare with (e). Bias voltage: (a,c) -4.0 V; (b) -4.0 V; (e) 0.9 V; (d, f) 1.0 V.

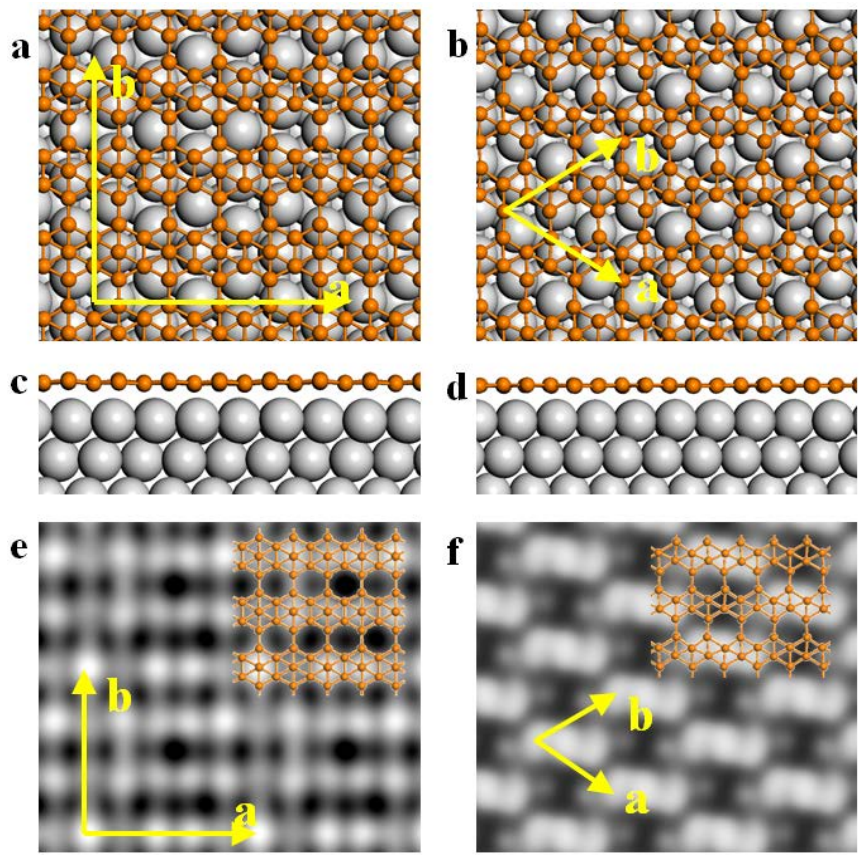

Fig.2: **Structure models of 2D boron sheets.** (a,b) Top view of the structure models of S1 and S2, respectively. (c,d) Side view of the structure models of S1 and S2, respectively. The orange and gray balls represent the B and Ag atoms, respectively. (e,f) Simulated STM topographic images of S1 and S2 phases, respectively. The basic vectors of unit cell are marked by yellow arrows.

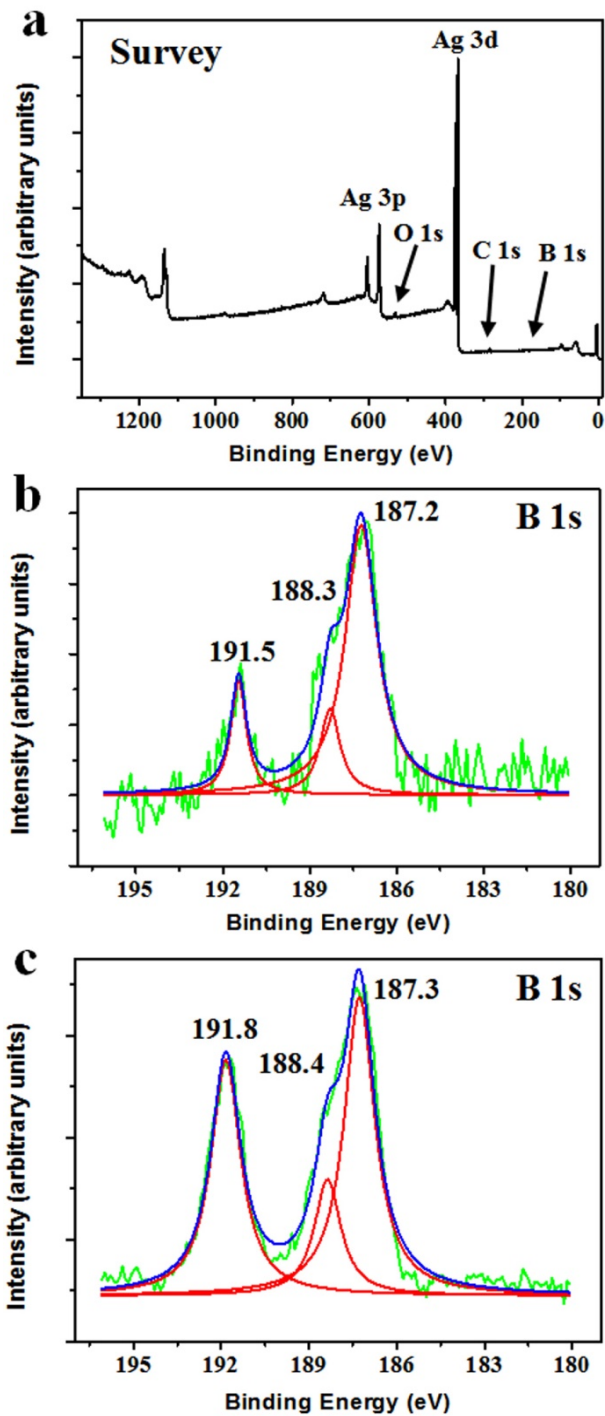

Fig. 3: **XPS of 2D boron sheets on Ag(111).** (a, b) Full scale survey spectrum and B 1s spectrum of boron sheets on Ag(111) (S1 phase) with B coverage 0.7 ML. (c). B 1s spectrum of boron sheets on Ag(111) with B coverage 1.0 ML. The peaks were fitted using combined Gaussian-Lorentzian function. The ratio of Lorentzian function is fitted to be (b) 93%, 81%, 100%; (c) 100%, 100%, 99%, respectively. The green, red and blue curves correspond to original data, fitting lines, sum of fitting lines, respectively.

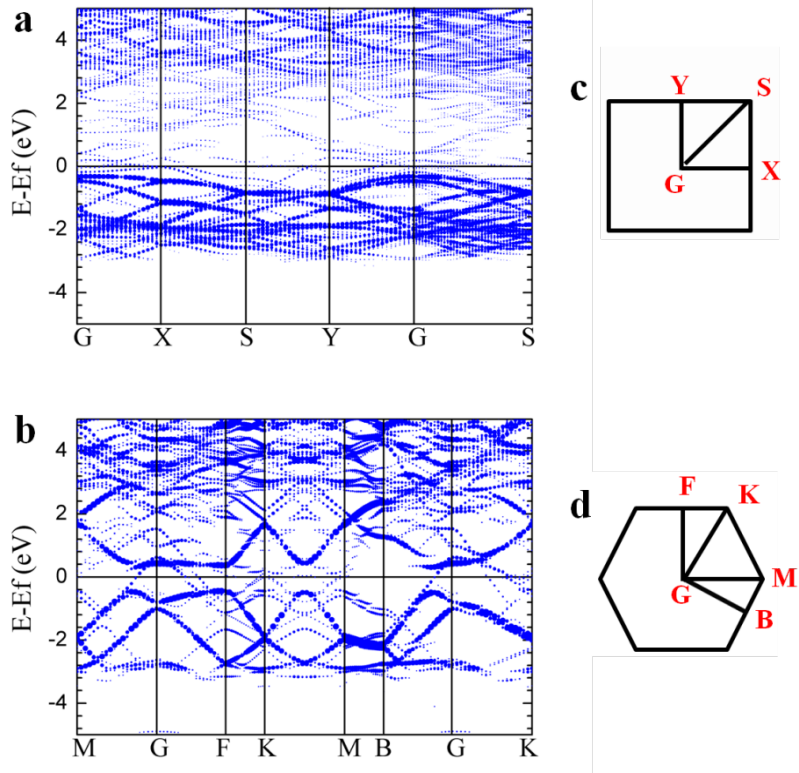

Fig.4: **Projected band structures of 2D boron sheets on Ag(111).** (a, b) Band structures of S1 and S2, respectively. The size of blue dot denotes to the contribution weight. (c, d) The high symmetric points of the 2D Brillouin zones for S1 and S2, respectively.